\documentclass[12pt]{article}
\usepackage{latexsym}
\usepackage{epsfig}

\usepackage{graphicx}
\usepackage{epstopdf}

\usepackage{epsfig,amssymb,amsmath,amscd}

\newcommand{\bq}{\begin{equation}}
\newcommand{\eq}{\end{equation}}
\newcommand{\bqa}{\begin{eqnarray}}
\newcommand{\eqa}{\end{eqnarray}}
\newcommand{\ben}{\begin{enumerate}}
\newcommand{\een}{\end{enumerate}}
\newcommand{\bc}{\begin{center}}
\newcommand{\ec}{\end{center}}
\newcommand{\bqb}{\begin{eqnarray*}}
\newcommand{\eqb}{\end{eqnarray*}}

\def\pr#1#2#3{Phys. Rev. ${\bf{#1}}$, #2 (#3)}
\def\prl#1#2#3{Phys. Rev. Lett. ${\bf{#1}}$, #2 (#3)}

\def\epj#1#2#3{Eur. Phys. J. ${\bf{#1}}$, #2 (#3)}

\def\jmp#1#2#3{J. Mod. Phys. ${\bf{#1}}$, #2 (#3)}

\begin{document}
\pagenumbering{arabic}
\thispagestyle{empty}
\def\thefootnote{\fnsymbol{footnote}}
\setcounter{footnote}{1}

\begin{flushright}
Jan.31, 2017\\
 \end{flushright}

\begin{center}
{\Large {\bf Test of Higgs boson compositeness\\
in $ZH$ production through gluon-gluon and photon-photon collisions}}.\\
 \vspace{1cm}
{\large F.M. Renard}\\
\vspace{0.2cm}
Laboratoire Univers et Particules de Montpellier,
UMR 5299\\
Universit\'{e} Montpellier II, Place Eug\`{e}ne Bataillon CC072\\
 F-34095 Montpellier Cedex 5, France.\\
\end{center}

\vspace*{1.cm}
\begin{center}
{\bf Abstract}
\end{center}

In the spirit of the Composite Standard Model (CSM) we look at the
effect of $H$, $Z_L$ compositeness on $ZH$ production in gluon-gluon 
and photon-photon collisions. 
Because they would perturb the basic cancellations occuring in SM 
amplitudes, such compositeness effects would lead to spectacular 
modifications of the energy dependence of the amplitudes 
and cross sections. We illustrate this fact by introducing
$H$, $Z_L$ test form factors and computing the corresponding differential 
cross section and the fraction of longitudinal $Z_L$ production.

\vspace{0.5cm}

\def\thefootnote{\arabic{footnote}}
\setcounter{footnote}{0}
\clearpage

\section{INTRODUCTION}

We have recently considered the possibility of Higgs boson compositeness
in a way which preserves the main features of the SM at low energies,
in particular those resulting from gauge invariance and Goldstone equivalence;
we called it the Composite Standard Model picture (CSM), \cite{CSM}.\\
The motivation for Higgs boson compositeness can be found in ref.\cite{Hcomp}
and (possibly for other sectors) in \cite{comp}, \cite{Portal}, 
\cite{BSMth}. We had already looked for several tests of Higgs boson compositeness
and of fact that the Higgs boson could be a portal to new sectors (possibly
involving unvisible states), see ref.\cite{Hincl,HHH,Htt}.\\
In the CSM spirit, compositeness of the Higgs sector would affect in a
similar way all members of the Higgs doublet, i.e. $H$, $G^{\pm}$ and $G^0$
and through equivalence, the $W_L$, $Z_L$ states. A form factor
(similarly to the hadronic case) with a new physics scale $M$ may affect
the $H$, $G^{\pm}$ and $G^0$ couplings.
Such a form factor should be close to 1 at low $q^2$ but, after showing some structures
around the new physics scale ($q^2\simeq M^ 2$), it may decrease at very high  $q^2$.
In a previous paper \cite{CSM} we have shown that the process $e^+e^-\to ZH$
(see \cite{ZH1,ZH2}) is particularly adequate to the study of such a form factor and will even 
furnish the basic source of an input for predicting CSM effects in other processes.\\

In the present paper we consider the processes
of $ZH$ production through gluon-gluon and photon-photon collisions.
These processes occur at one loop but present a peculiar feature,
a cancellation of the contributions of triangle and box diagrams in SM, which 
could be perturbed by the above mentioned compositeness effects. 
We will show that, indeed, these
processes should be particularly sensitive to the presence of form
factors in the $H$ and $Z_L$ couplings.\\
We will show how the energy and the angular dependences of the cross sections
are immediately modified by the presence of a form factor, even when
the new physics scale is rather high.\\
We also show that, in SM and in CSM, the percentage of $Z_L$ polarization 
in these $ZH$ processes is always very large, this fact allowing a clean
study of the $Z_L$ state.\\
We finally give the result of the addition of similar very heavy quark loops
to the SM quark loops (triangle and box). This shows again the large sensitivity
of these processes to new physics effects.\\

Contents: Section 2 is devoted to the detailed study of the SM amplitudes
and cross section of the $gg\to ZH$ process; Section 3 gives the effects
of the presence of form factors in the $G^0Z_LH$ and $ZZ_LH$ couplings
and of the addition of a very heavy new quark loop. 
Summary and conclusions are given in Section 3.\\

\section{THE $gg\to ZH$ PROCESS}

At one loop the $gg\to ZH$ process is described by 3 types of diagrams
shown in Fig.1a,b,c:\\
 (1a) an initial quark triangle connected to an
intermediate $G^0$ goldstone boson coupled to $ZH$,\\
 (1b) a similar one with an
intermediate $Z$ boson and\\
 (1c) a set of (direct, crossed and twisted) quark boxes.\\

With usual SM couplings the resulting helicity amplitudes $F_{\lambda,\lambda',\tau}$
for $\lambda,\lambda' =\pm{1\over2}$ and $\tau=0,\pm1$,
can be explicitly obtained from the $\gamma\gamma\to ZH$ case given in eq.(B26) 
and (B29-35) of \cite{gagaVH}, by just replacing the photon quark coupling $(-eQ_q)$
by the gluon one ($g_s{\lambda^i\over2}$), see \cite{ggVH}.
CP conservation impose the relations 
\bq
F_{---}=F_{+++} ~~~~F_{--0}=-F_{++0}~~~~ F_{--+}=F_{++-} 
\eq
\bq
F_{-+-}=F_{+-+}~~~~  F_{-+0}=-F_{+-0}~~~~F_{-++}=F_{+--} 
\eq
Bose statistics imposes in addition the angular constraint
\bq
F_{\lambda,\lambda',\tau}(\theta)=(-1)^{\tau}F_{\lambda',\lambda,\tau}(\pi-\theta)
\eq

The resulting unpolarized $gg\to ZH$ differential cross section is then obtained 
\bq
{d\sigma\over d\cos\theta}={p\over512\pi s\sqrt{s}}\sum_{\lambda,\lambda',\tau} 
|F_{\lambda,\lambda',\tau}|^2
\eq
with 
\bq
p={\sqrt{(s-(m_H+m_Z)^2)(s-(m_H-m_Z)^2)}\over2\sqrt{s}}
\eq
One will also consider the purely longitudinal $Z_L$ production
\bq
{d\sigma_L\over d\cos\theta}={p\over512\pi s\sqrt{s}}\sum_{\lambda,\lambda',\tau=0} 
|F_{\lambda,\lambda',\tau}|^2
\eq
and the fraction
\bq
P(Z_L)={d\sigma_L\over d\cos\theta}/{d\sigma\over d\cos\theta}
\eq
It is now important to see how each diagram contributes and what is the result of their combination.
The diagrams (1a,b) only contributes to the $F_{\pm\pm0}$ amplitudes whereas the diagrams
(1c) contribute to all amplitudes. The results for the real and imaginary parts of the 
amplitudes (at $\theta={\pi\over3}$) are shown in Fig.2,3.
The remarkable high energy features are the fact that the triangle
(1a) and the box (1c) contributions to $F_{\pm\pm0}$ almost cancel  and that
the (1b) contribution is much weaker. This can be checked by using the high energy
expansions of the expressions of the amplitudes in terms of Passarino-Veltman
functions (see \cite{gagaVH}, \cite{ggVH}). \\
The (1a) contribution is
\bq
F_{\pm\pm0}=\mp~{\alpha\alpha_s psm^2_t\sqrt{s}\over s^2_Wc^2_Wm^2_Z(s-m^2_Z)}C^t_0(s)
\eq
At high energy the leading expression is given by
\bq
C^t_0(s) \to {1\over2s}ln^2{s\over m^2_t}
\eq
and it is cancelled by an opposite term coming from the (1c) contribution resulting from
the addition of the 3 types of box terms.\\
This behaviour can be seen in Fig.2,3 where one sees (especially for its 
imaginary part) that the high value of the $F_{\pm\pm0}$ amplitudes at low energy is
progressively strongly reduced at high energy so that these amplitudes become 
comparable or smaller than the other ones. Finally the $F_{\pm\mp 0}$ amplitudes are
the dominant ones (mainly from their imaginary part).\\
These results agree with the helicity conservation (HC) rule \cite{hc} which predicts 
the dominance of the $F_{\pm\mp 0}$ amplitudes
(exact in MSSM, but only approximate in SM).\\

The corresponding energy dependence, the angular distribution (illustrated
at 1 and 4 TeV) and the fraction of $Z_L$ production, for the 
SM cross section are shown in Fig.4-7 together with the ones resulting 
from the CSM cases that we will present in the next section.\\
After a strong threshold peak (of the top triangles and boxes) around $2m_t$, 
the SM cross section quickly decreases with the energy. 
The (symmetrical) angular distribution is rather flat and the $Z_L$
fraction is important, varying between 80 and 100 percent.\\

All the above features also occur in $\gamma\gamma\to ZH$. Apart from
small leptons and light quarks contributions and the adequate change in the global
normalization (replacing gluon by photon couplings), formulas for amplitudes and 
cross sections are the same.\\

\section{A Compositeness Form Factor}

We now want to look for compositeness signals of the Higgs sector
in the CSM spirit. This assumes that the basic SM structure is maintained at low energy,
no anomalous couplings is introduced. At high energy,
composite Higgs and Goldstone bosons couplings will only be affected 
by form factors revealing their composite nature.\\

For the considered $gg\to ZH$ process with the diagrams of Fig.1
we will first consider the effect of form factors occuring in $G^0Z_LH$ 
and $ZZ_LH$ couplings.\\

For the illustrations we will use the following expression
\bq
F_H(s)={(m_Z+m_H)^2+M^2\over s+M^2}
 \label{FF}
\eq
where $M$ is a new physics scale (taking the value of $2$ or $5$ TeV in the illustrations).

The diagrams (1a) and (1b) are directly affected but not
the (1c) box. This has an immediate consequence for the cancellation between (1a) and (1c)
mentioned in the previous Section for the SM case. The reduction 
of (1a), whereas (1c) is not affected, immediately leads to an increase
of the resulting $F_{\pm\pm0}$ amplitudes.\\

The consequences for the observables are illustrated in Fig.4 for the energy dependence
of the cross section, in Fig.5,6 for the angular distribution and in Fig.7 for
the $Z_L$ fraction. One can see that, even with a high mass scale of 5 TeV, 
the form factor leads to a strong observable effect.\\

{\bf Effect of Additional Loops}\\

For comparison we now just want to see what could be the effect
of additional loops connecting gluons and a composite Higgs sector.
For simplicity we consider the loops due to a heavy quark (its mass $M_Q$ is taken as 
2 or 5 TeV in the illustrations) with couplings to gluons, $Z$ and $H$
similar to those of the top quark.\\
This arbitrary chloice could correspond to many different possibilities, a new heavy quark 
sector, Higgs sector constituents, .... .\\
The resulting modifications of the energy and angular distributions
and of the $Z_L$ fraction can be seen in Fig.4-7.\\

For the same reasons as in the case of the simple form factor, the basic SM cancellations
being perturbed, an increase of the cross section is immediately generated with 
threshold effects around $2M_Q$, as one can see in Fig.4.\\

Obviously, if departures to SM predictions are observed one day, detailed analyses
of many different processes would be required in order to identify their origin.\\

\section{CONCLUSIONS}

Summarizing, we have shown that the processes $gg\to ZH$ and 
$\gamma\gamma\to ZH$ would be strongly sensitive
to, even very small, Higgs compositeness effects, for example those occuring
in the CSM.\\
We base this conclusion on the computation of the one loop amplitudes and
cross sections including the modifications generated by the presence of form factors
in $G^0Z_LH$ and $ZZ_LH$ couplings. Because of strong cancellations occuring
among SM contributions, the smallest modification, for example due to the above mentioned
form factor, will immediately produce a very large observable effect.\\
This is the basic feature of the considered one loop processes.\\ 
This type of strong effect is very specific of this peculiar breaking of the SM cancellation.
It is stronger than what would appear with the form factor effect
at Born level in the $e^+e^-\to ZH$ process and in the one loop  $gg\to HH$ and 
$\gamma\gamma\to HH$ processes which do not involve this peculiar SM cancellation. 
It does also not appear in other one loop Higgs production processes, for
example in the simple  $gg\to H$ or $\gamma\gamma\to H$ processes.\\
These strong modifications of the cross section keep the angular distribution rather flat and also 
the high value of the fraction of longitudinal $Z_L$ production (apart from very local
structures appearing when the lack of cancellation is maximal). These features should help
for the identification of the process.\\
We beleive that it will be worthwhile to perform detailed experimental
and phenomenological studies of these processes, first in hadronic
collisions and possibly at future $\gamma\gamma$ colliders \cite{gammagamma}.

\newpage

\begin{figure}[p]
\[
\epsfig{file=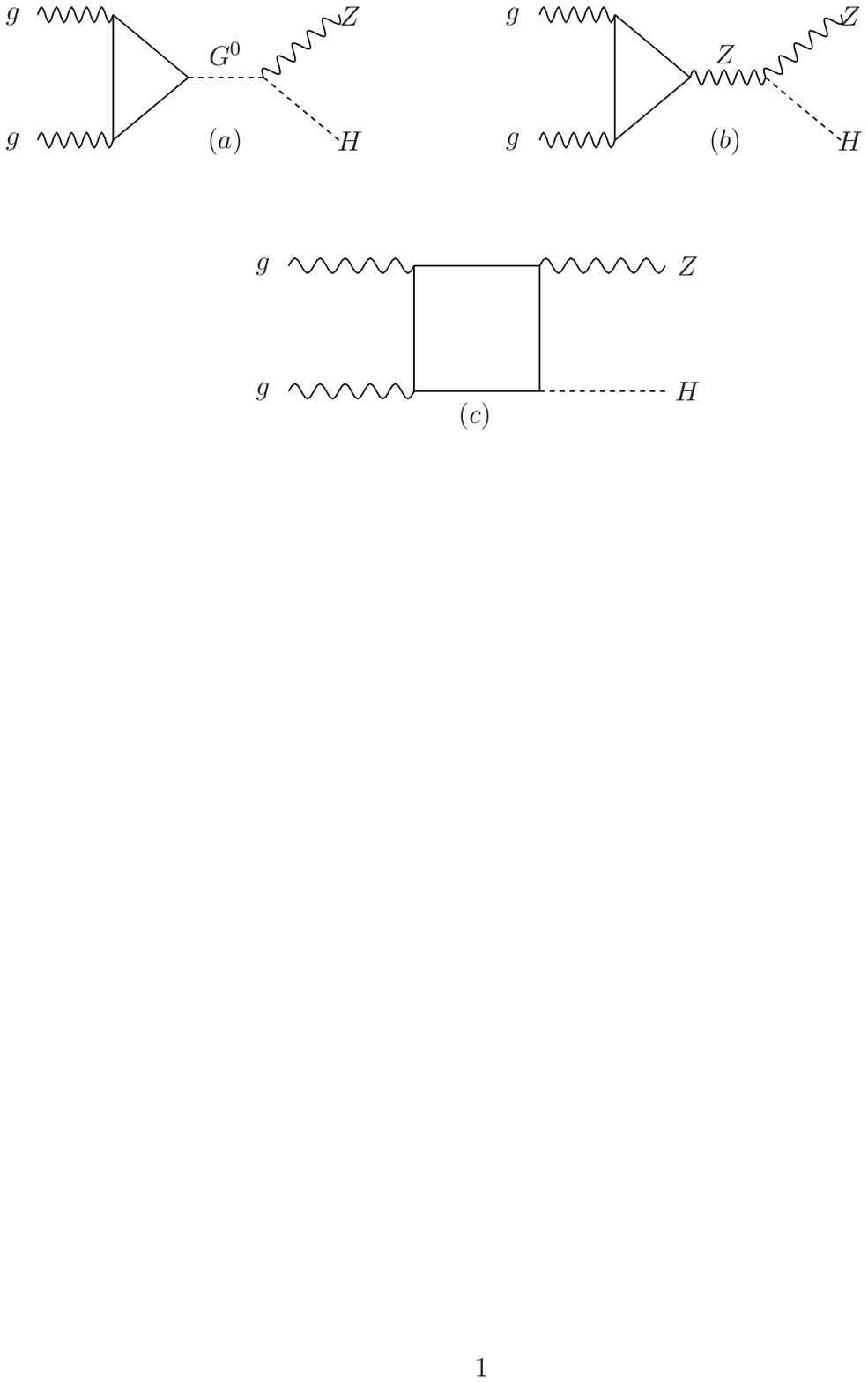, height=24.cm}
\]\\
\vspace{-15cm}
\caption[1] {The 3 classes of one loop diagrams contributing
to the $gg\to ZH$ process; (c) involving also the twisted box.}
\end{figure}

\clearpage

\begin{figure}[p]
\[
\epsfig{file=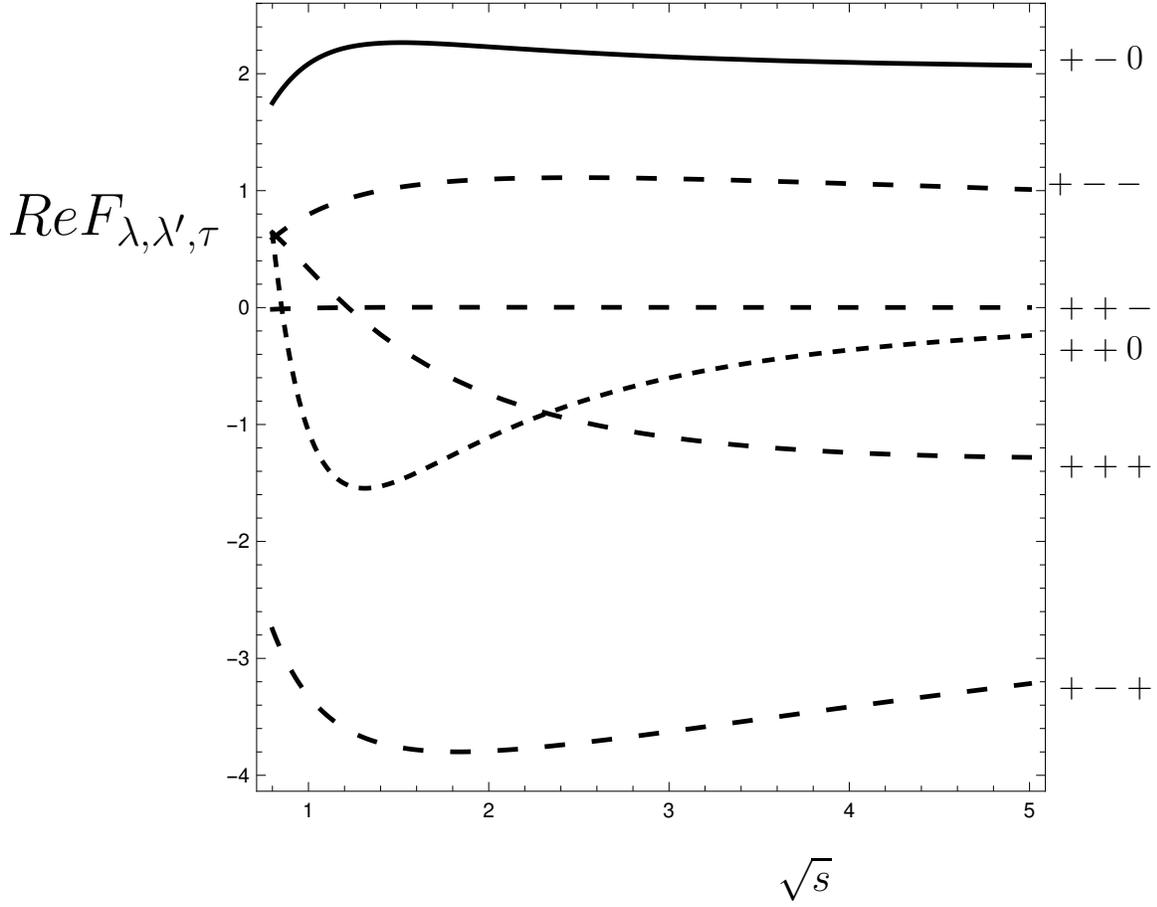, height=12.cm}
\]\\
\vspace{-1cm}
\caption[1] {Real parts of the 6 independent SM amplitudes.}
\end{figure}

\clearpage

\begin{figure}[p]
\[
\epsfig{file=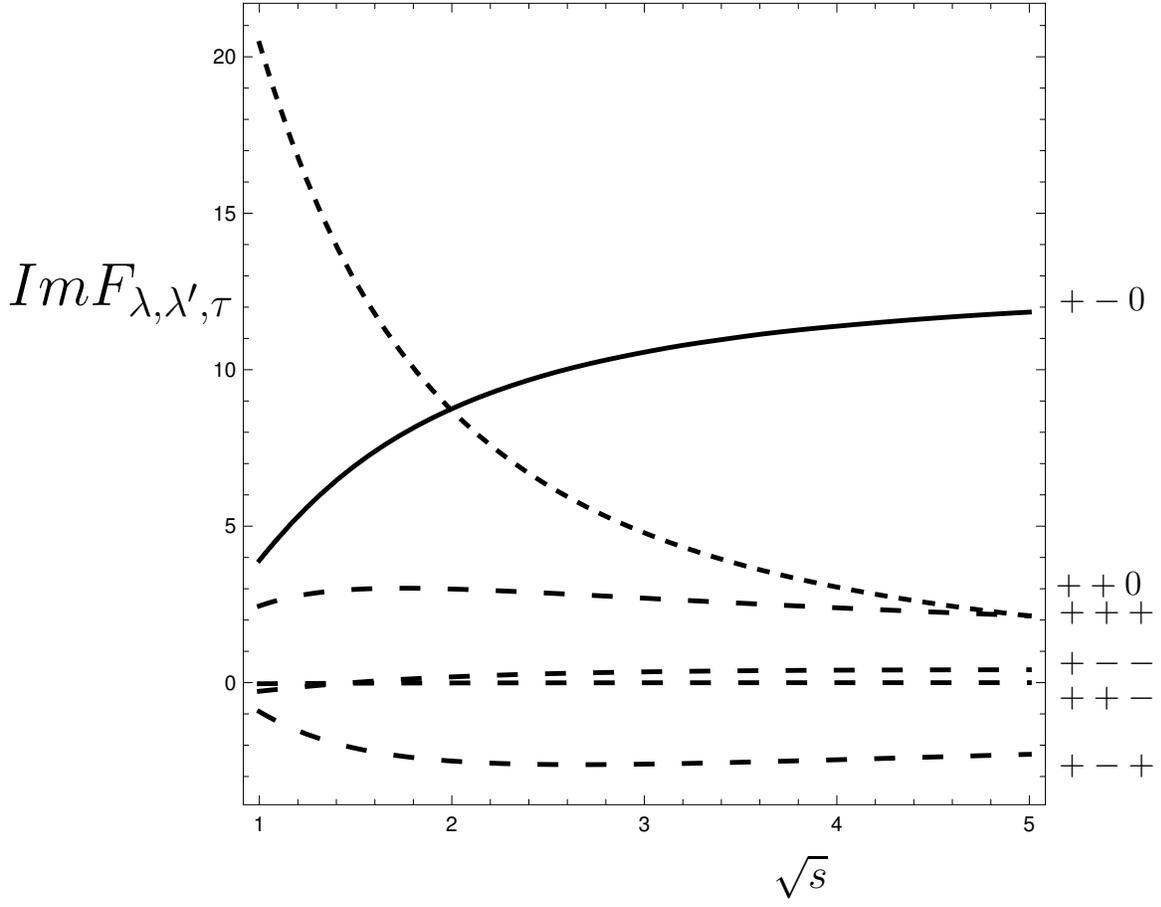, height=12.cm}
\]\\
\vspace{-1cm}
\caption[1] {Imaginary  parts of the 6 independent SM amplitudes.}
\end{figure}

\clearpage

\begin{figure}[p]
\[
\epsfig{file=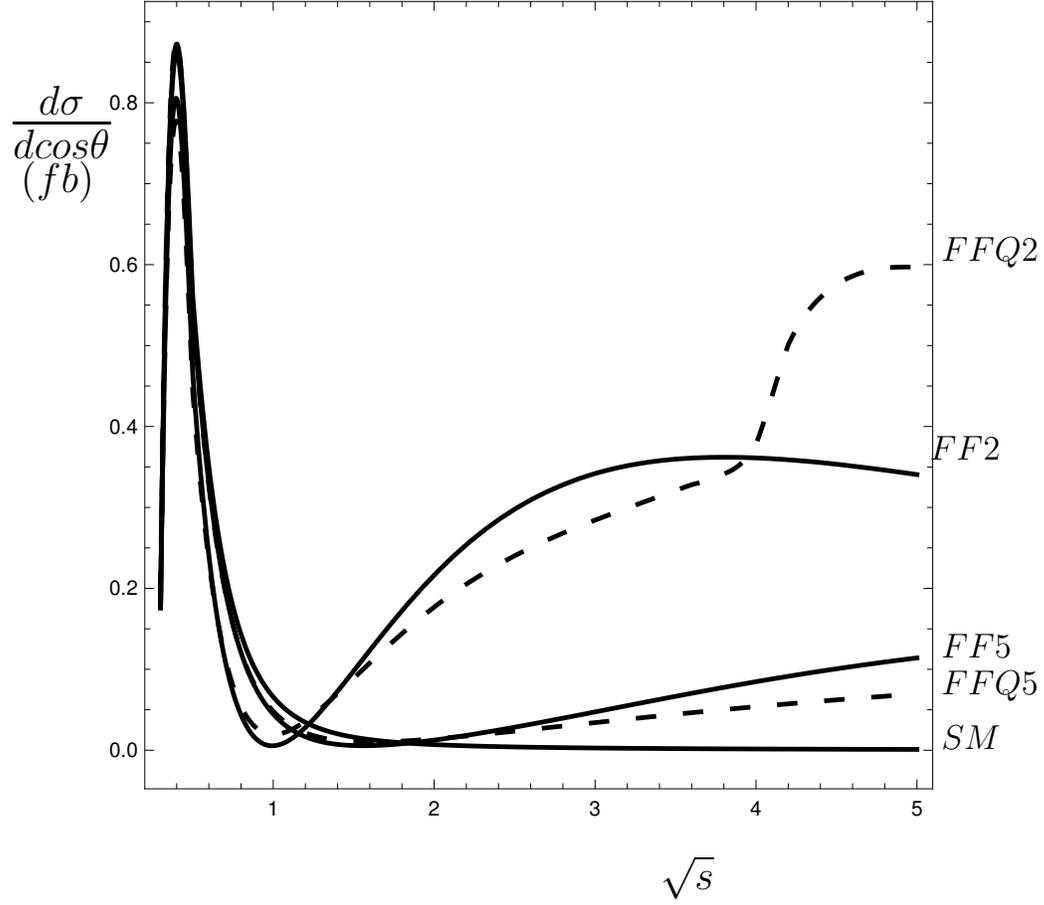, height=12.cm}
\]\\
\vspace{-1cm}
\caption[1] {Energy dependence of the differential cross section at $\theta={\pi\over3}$; 
in SM; with a form factor and a mass scale of 2 or 5 TeV (FF2, FF5); with, in addition, 
a heavy quark loop with a mass of of 2 or 5 TeV (FFQ2, FFQ5).}
\end{figure}

\clearpage

\begin{figure}[p]
\[
\epsfig{file=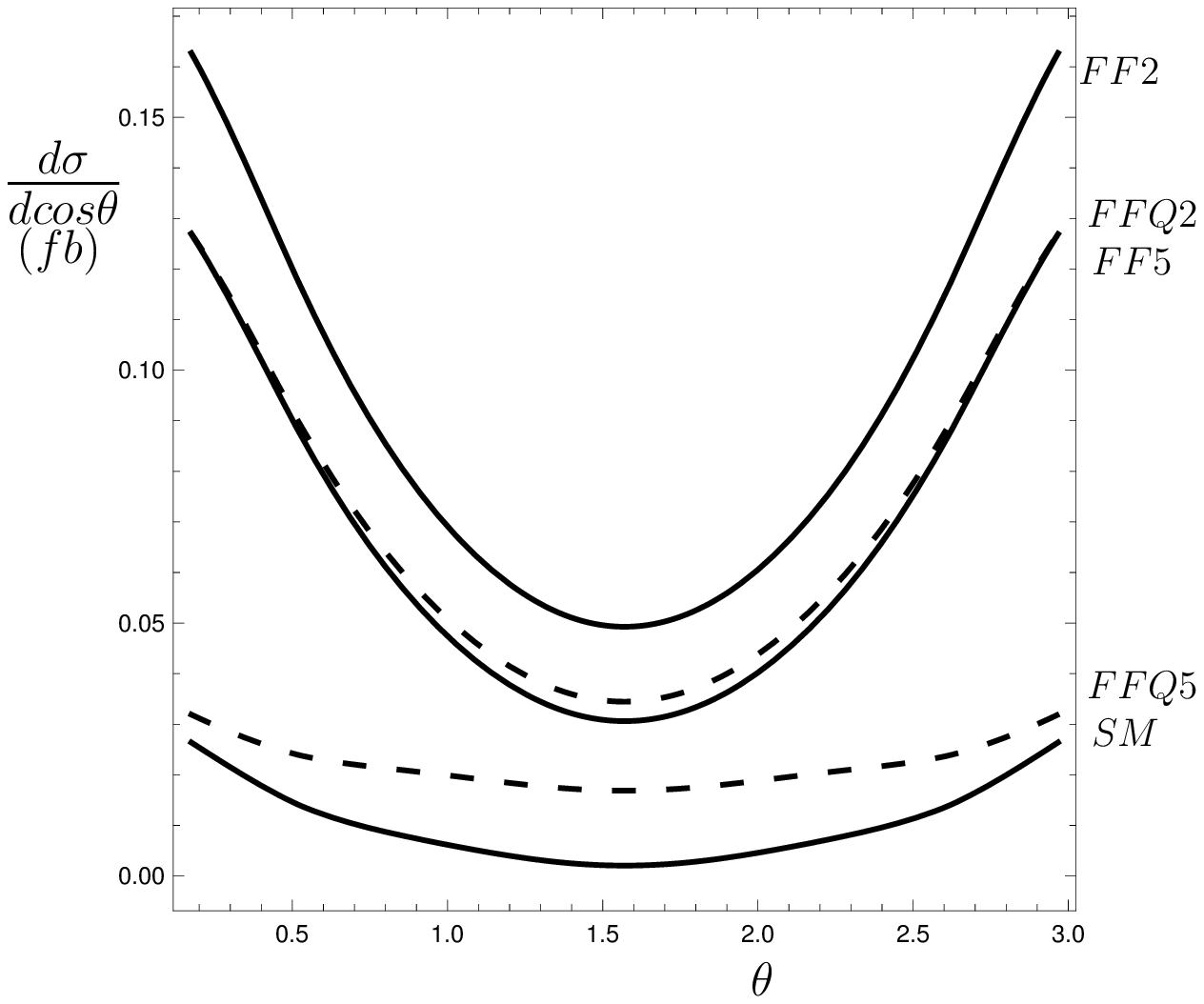, height=12.cm}
\]\\
\vspace{-1cm}
\caption[1] {Angular dependence of the differential cross section at $\sqrt{s}=1$ TeV;
same notations as in Fig.4.}
\end{figure}

\clearpage

\begin{figure}[p]
\[
\epsfig{file=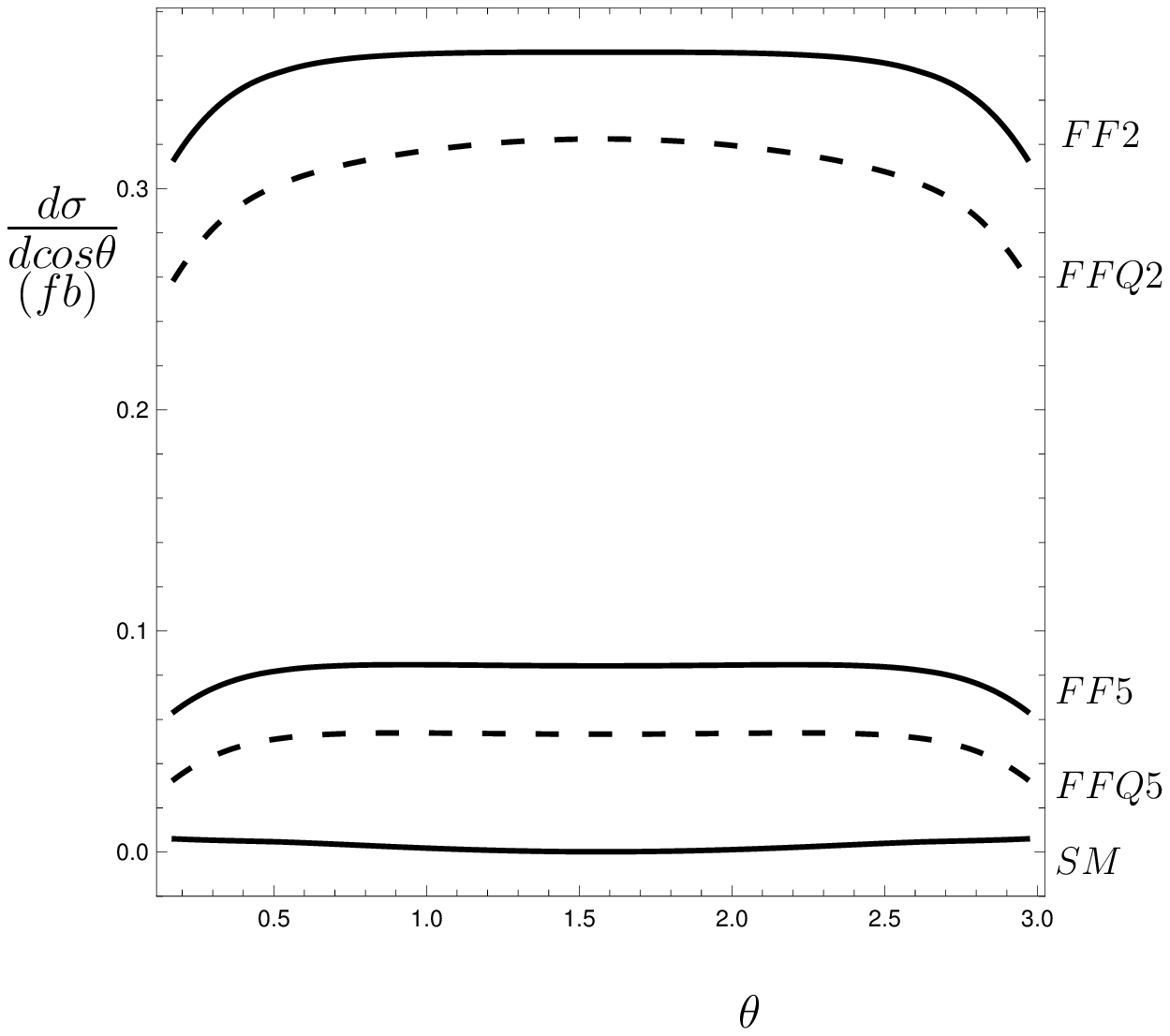, height=12.cm}
\]\\
\vspace{-1cm}
\caption[1] {Angular dependence of the differential cross section at $\sqrt{s}=4$ TeV;
same notations as in Fig.4.}
\end{figure}

\clearpage

\begin{figure}[p]
\[
\epsfig{file=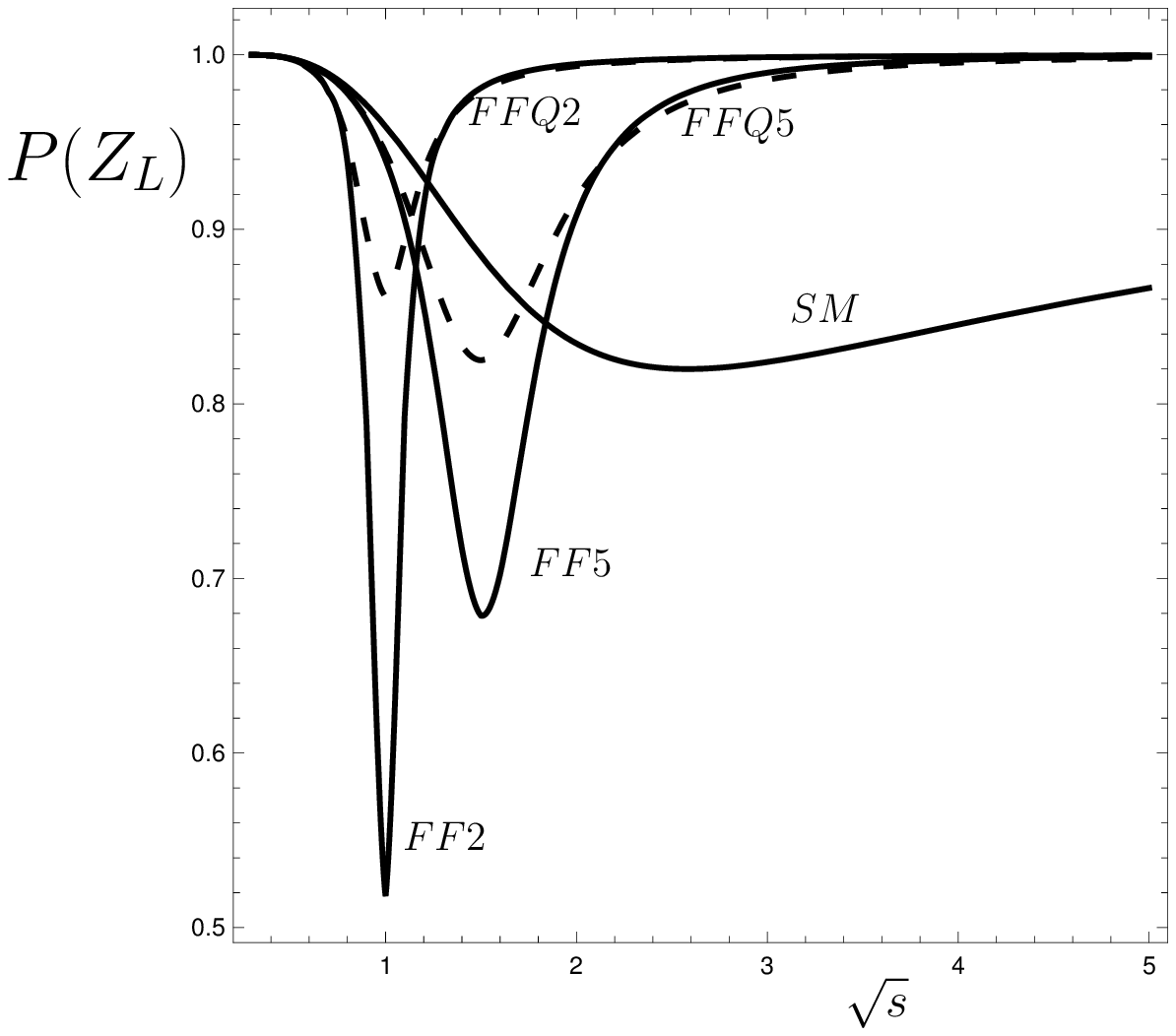, height=12.cm}
\]\\
\vspace{-1cm}
\caption[1] {Energy dependence of the fraction of longitudinal $Z_L$ production
for $\theta={\pi\over3}$; same notations as in Fig.4.}
\end{figure}

\clearpage


\clearpage

\begin{thebibliography}{99}

%
\bibitem{CSM} F.M. Renard,
arXiv: 1701.04571.
%
\bibitem{Hcomp} G. Panico and A. Wulzer, Lect.Notes Phys. {\bf 913},1(2016).

%
\bibitem{comp}  H. Terazawa, Y. Chikashige and K. Akama, \pr{D15}{480}{1977};
for other references see
H. Terazawa and M. Yasue, Nonlin.Phenom.Complex Syst. {\bf19},1(2016);
\jmp{5}{205}{2014}.
%
\bibitem{Portal} B.Patt and F. Wilczek, arXiv: hep-ph/0605188.
%
\bibitem{BSMth}
M.E. Peskin, Ann.Phys.(N.Y.){\bf 528},20(2016). M. Muhlleitner, arXiv:1410.5093. Ben Gripaios,
arXiv:1503.02636, arXiv:1506.05039.
%
\bibitem{Hincl}  G.J. Gounaris and F.M. Renard,
\pr{D92}{053011}{2015}.
%
\bibitem{HHH}  G.J. Gounaris and F.M. Renard,
\pr{D93}{093018}{2016}.
%
\bibitem{Htt}  G.J. Gounaris and F.M. Renard,
\pr{D94}{053009}{2016}.
%
\bibitem{ZH1} A. Djouadi, arXiv:1505.01059,1511.07853.
%
\bibitem{ZH2} S. Lukic, arXiv:1610.00628.
%
\bibitem{gagaVH}  G.J. Gounaris, P.I. Porfyriadis and F.M. Renard,
\epj{C20}{659}{2001};
%
\bibitem{ggVH}   G.J. Gounaris, J. Layssac and F.M. Renard,
\pr{D80}{013009}{2009}.
%
\bibitem{hc}  G.J. Gounaris and F.M. Renard,
\prl{94}{131601}{2005},  hep-ph/0501046; \pr{D73}{097301}{2006} ,  hep-ph/0604041.
%
\bibitem{gammagamma} V.I. Telnov, Nucl.Part.Phys.Proc. {\bf 273}(2016)219.







 
\end{thebibliography}
\end{document}